\documentclass[11pt,twoside]{stm}

\begin{document}
\markboth{\sc E.~Jur\v{c}i\v{s}inov\'a, M.~Jur\v{c}i\v{s}in, R.
Remecky, M. Scholtz}{\sc Anomalous Scaling of a Passive Scalar
Field}

\STM

\title{Influence of anisotropy and compressibility on anomalous scaling of a passive scalar field}

\authors{E.~Jur\v{c}i\v{s}inov\'a$^{1,2}$, M.~Jur\v{c}i\v{s}in$^{1,3}$, R.~Remecky$^4$, M.~Scholtz$^4$}

\address{$^{1}$\,Institute of Experimental Physics SAS, Watsonova 47, 04001 Ko\v{s}ice, Slovakia,\\
$^{2}$\,Laboratory of Information Technologies JINR, 141 980 Dubna,
Russia,\\$^{3}$\,Laboratory of Theoretical Physics JINR, 141 980
Dubna, Russia,
\\$^4$\,Department of Physics and Astrophysics, Institute of Physics,
P.J. \v{S}af\'arik University, \\ Park Angelinum 9, 04001
Ko\v{s}ice, Slovakia}
\bigskip

\begin{abstract}
Influence of uniaxial small-scale anisotropy and compressibility on
the stability of scaling regime and on the anomalous scaling of
structure functions of a scalar field is investigated in the model
of a passive scalar field advected by the compressible Gaussian
strongly anisotropic velocity field with the covariance $\propto
\delta(t-t^{\prime})|{\bf x}-{\bf x^{\prime}}|^{2\varepsilon}$ by
using the field theoretic renormalization group and the operator
product expansion. The inertial-range stability of the corresponding
scaling regime is established. The anomalous scaling of the
single-time structure functions is studied and the corresponding
anomalous exponents are calculated. Their dependence on the
compressibility parameter and anisotropy parameters is analyzed. It
is shown that the presence of compressibility leads to the
decreasing of the critical dimensions of the important composite
operators, i.e., the anomalous scaling is more pronounced in the
compressible systems. This result is demonstrated for the structure
function of the third order. All calculations are done to the first
order in $\varepsilon$.

\end{abstract}

\section*{Introduction}

The so-called "rapid change model" of a scalar field passively
advected by a self-similar Gaussian $\delta-$correlated in time
velocity field introduced by Kraichnan \cite{Kra68} and number of
its extensions play the central role in the theoretical
investigation of intermittency and anomalous scaling
\cite{MonYag75,Frisch95}. The main reason for this is the
experimental fact that the deviations from the statements of the
famous classical Kolmogorov-Obukhov phenomenological theory (see,
e.g., \cite{MonYag75,Frisch95}) are more noticeable for simpler
models of passively advected scalar quantity (scalar field) than for
the velocity field itself (see \cite{FaGaVe01} and references cited
therein) and, at the same time, the problem of the passive advection
of a scalar field is much easier from theoretical point of view than
the original problem of anomalous scaling in the framework of the
Navier-Stokes velocity field. Within the rapid change model of a
passive scalar advection the systematic analysis of the
corresponding anomalous exponents was done for the first time on the
microscopic level. For example, in the so-called "zero-mode
approach" to the rapid change model (see survey paper
\cite{FaGaVe01} and references cited therein) the anomalous
exponents are found from the homogenous solutions (zero modes) of
the closed equations for the single-time correlations.

One of the most effective approach for studying self-similar scaling
behavior is the method of the field theoretic renormalization group
(RG) \cite{ZinnJustin,Vasiliev}. This method can be also used in the
theory of fully developed turbulence and related problems, e.g., in
the problem of a passive scalar advection by the turbulent
environment (see \cite{Vasiliev,AdAnVa96,AdAnVa99} and references
cited therein).

In \cite{AdAnVa98} the field theoretic RG and operator-product
expansion (OPE) were used in the systematic investigation of the
rapid-change model of a passive scalar. It was shown that within the
field theoretic approach the anomalous scaling is related to the
very existence of the so-called "dangerous" composite operators with
negative critical dimensions in the OPE (see, e.g.,
Refs.\,\cite{Vasiliev,AdAnVa99} for details).

In subsequent papers a few generalizations of the rapid-change model
towards more realistic ones were done.  For example, in
\cite{AdAnHnNo00} the field theoretic RG and OPE were applied to the
rapid change model of a passive scalar advected by the
$\delta$-correlated, Gaussian strongly anisotropic velocity field.
It was shown that in the small-scale anisotropic case the anomalous
exponents of the structure functions and correlation functions are
nonuniversal and they are functions of the parameters of anisotropy,
Besides, they form the hierarchy with the leading exponent related
to the most "isotropic" operator. On the other hand, in
\cite{AnHo01} the influence of compressibility and large-scale
anisotropy on the anomalous scaling behavior was studied in the
aforementioned model and the anomalous exponents of higher-order
correlation functions were calculated as functions of the parameter
of compressibility. It was shown that the presence of small-scale
anisotropy is more pronounced for larger values of the
compressibility parameter $\alpha$.  From this point of view,
compressible systems are very interesting to be studied.

In the end, in \cite{JuJuReSc06a} the combined effects of the
small-scale anisotropy and compressibility on the anomalous scaling
of the structure functions of a passive scalar within the
rapid-change model were studied by using the simplest possible
generalization of the incompressible and anisotropic tensor
structure of the velocity field correlator to the compressible one
where compressibility was introduced only to the isotropic part of
the anisotropic tensor structure. In present paper, we shall
investigate more general case of the model where compressibility
will be introduced through all anisotropic tensor structures of the
velocity field statistics. First of all, we shall establish
stability of the scaling regime of the model and coordinates of the
corresponding infrared (IR) stable fixed point will be found as
functions of the compressibility and anisotropy parameters. These
results will be then used in the analysis of the asymptotic behavior
of the single-time structure functions of a passively advected
scalar field.

\section*{Formulation of the model}

We shall investigate the model of the advection of a passive
"tracer" $\theta(x)\equiv \theta(t,{\bf x})$ which is described by
the following stochastic equation
\begin{equation}
\partial_t \theta = \nu_0 \triangle \theta - (v_i
\partial_i) \theta + f, \label{stoch1}
\end{equation}
where $\partial_t \equiv \partial/\partial t$, $\partial_i\equiv
\partial/\partial x_i$, $\triangle\equiv \partial^2$ is
the Laplace operator, $\nu_0$ is the coefficient of molecular
diffusivity (a subscript $0$ will denote bare parameters of
unrenormalized theory), $v_i \equiv v_i(x)$ is the $i$-th component
of the compressible velocity field ${\bf v}(x)$, and $f \equiv f(x)$
is a Gaussian random noise with zero mean and correlation function
\begin{equation}
D^f \equiv\langle f(x) f(x^{\prime})\rangle =
\delta(t-t^{\prime})C({\bf r}/L), \,\,\, {\bf r}={\bf x}-{\bf
x^{\prime}},\label{correlator}
\end{equation}
where parentheses $\langle...\rangle$ hereafter denote average over
corresponding statistical ensemble. The concrete form of the noise
defined in (\ref{correlator}) will not be essential in what follows.
The only condition which must be satisfied by the function $C({\bf
r}/L)$ is that it must decrease rapidly for $r\equiv |{\bf r}| \gg
L$, where $L$ denotes an integral scale related to the stirring.

In real problems the velocity field ${\bf v}(x)$ satisfies
Navier-Stokes equation but, in what follows, we shall work with a
simplified model where we suppose that the velocity field obeys a
Gaussian statistics with zero mean and pair correlation function
\begin{equation}
\langle v_i(x) v_j(x^{\prime}) \rangle \equiv D^v_{ij}(x;
x^{\prime})=  D_0 \delta(t-t^{\prime}) \nonumber \\
\int \frac{d^d k}{(2\pi)^{d}} \frac{R_{ij}({\bf
k})}{(k^2+m^2)^{d/2+\varepsilon}} \exp[i{\bf k}({\bf x}-{\bf
x^{\prime}})], \label{corv}
\end{equation}
where $d$ is the dimension of the space, ${\bf k}$ is the wave
vector, and $D_0$ is an amplitude factor related to the coupling
constant $g_0$ of the model (expansion parameter in the perturbation
theory, see, e.g., \cite{JuJuReSc06a}) by the relation
$D_0/\nu_0\equiv g_0\equiv\Lambda^{2\varepsilon}$, where $\Lambda$
is the characteristic UV momentum scale.  The parameter of the
energy spectrum of the velocity field $0<\varepsilon<1$ is taken in
such a way that its "Kolmogorov" value (the value which corresponds
to the Kolmogorov scaling of the velocity correlation function in
developed turbulence) is $\varepsilon=2/3$, and $1/m$ is another
integral scale. In our uniaxial anisotropic and compressible case
the tensor $R_{ij}({\bf k})$ is taken in the following way
\cite{AdAnHnNo00,JuJuReSc06a}
\begin{equation}
R_{ij} ({\bf k}) = \left(1 + \alpha_{1} \xi_k^2\right)
(P_{ij}+\alpha Q_{ij})+ \alpha_{2} n_s n_l (P_{is}+\alpha Q_{is})
(P_{jl}+\alpha Q_{jl})\,, \label{Tg-ij}
\end{equation}
where we denote $\xi_k={\bf n}\cdot {\bf k}/k$,  $P_{ij} ({\bf
k})\equiv \delta_{ij}-k_i k_j/k^2$ is common isotropic transverse
projector, $Q_{ij}=k_i k_j/k^2$ is the longitudinal projector, the
unit vector ${\bf n}$ determines the distinguished direction of
uniaxial anisotropy, $\alpha \geq 0$ is a free parameter of
compressibility, and $\alpha_{1}$, $\alpha_{2}$ are parameters
characterizing the anisotropy. From the positiveness of the
correlation tensor $D^v_{ij}$ one finds restrictions on the values
of the above parameters, namely, $\alpha_{1,2}>-1$.

The stochastic problem (\ref{stoch1})-(\ref{corv}) can be rewritten
in a field theoretic form with the following action functional
\cite{ZinnJustin,Vasiliev}
\begin{eqnarray}
S(\Phi)&=&\int dt\,d^d{\bf x}\,\, \theta^{\prime}\left[-\partial_t -
v_i\partial_i+\nu_0\triangle + \chi_0 \nu_0 ({\bf n}\cdot{\bf
\partial})^2\right]\theta -\frac{1}{2} \int
dt_1\,d^d{\bf x_1}\,dt_2\,d^d{\bf x_2} \nonumber \\ &&
\hspace{-1.8cm}\left( v_i(t_1,{\bf x_1}) [D_{ij}^v(t_1,{\bf
x_1};t_2,{\bf x_2})]^{-1} v_j(t_2,{\bf x_2}) -
\theta^{\prime}(t_1,{\bf x_1}) D^f(t_1,{\bf x_1};t_2,{\bf x_2})
\theta^{\prime}(t_2,{\bf x_2})\right), \label{action2}
\end{eqnarray}
where $D_{ij}^v$ and $D^f_{ij}$ are given in (\ref{corv}) and
(\ref{correlator}) respectively, $\theta^{\prime}$ is an auxiliary
scalar field (see, e.g., \cite{Vasiliev}), and the required
summations over the vector indices are implied. In action
(\ref{action2}) the term with new parameter $\chi_{0}$ is related to
the presence of small-scale anisotropy and the introduction of this
term is necessary to make the model multiplicatively renormalizable.
Model (\ref{action2}) corresponds to a standard Feynman diagrammatic
technique (see, e.g., \cite{AdAnHnNo00,JuJuReSc06a} for details) and
the standard analysis of canonical dimensions shows which
one-irreducible Green functions can possess UV superficial
divergences. The functional formulation (\ref{action2}) gives
possibility to extract large-scale asymptotic behavior of the
correlation functions after an appropriate renormalization procedure
which is needed to remove the UV-divergences.

\section*{Influence of anisotropy and compressibility on the scaling regime of the model and
on the anomalous scaling}

\input epsf
   \begin{figure}[t]
     \vspace{-1cm}
       \begin{flushleft}
       \leavevmode
       \epsfxsize=7.5cm
       \epsffile{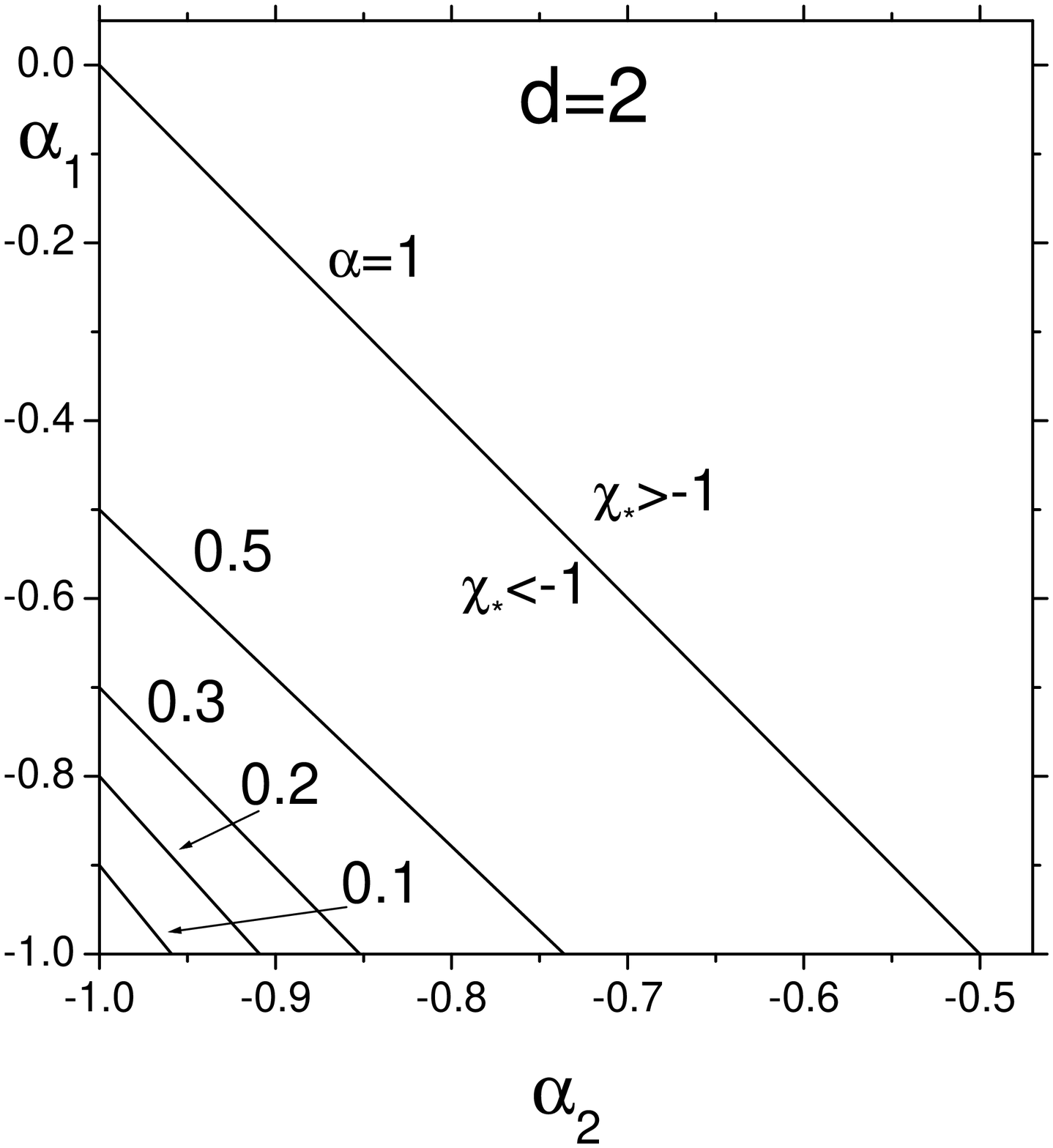}
   \end{flushleft}
     \vspace{-11.7cm}
   \begin{flushright}
       \leavevmode
       \epsfxsize=7.5cm
       \epsffile{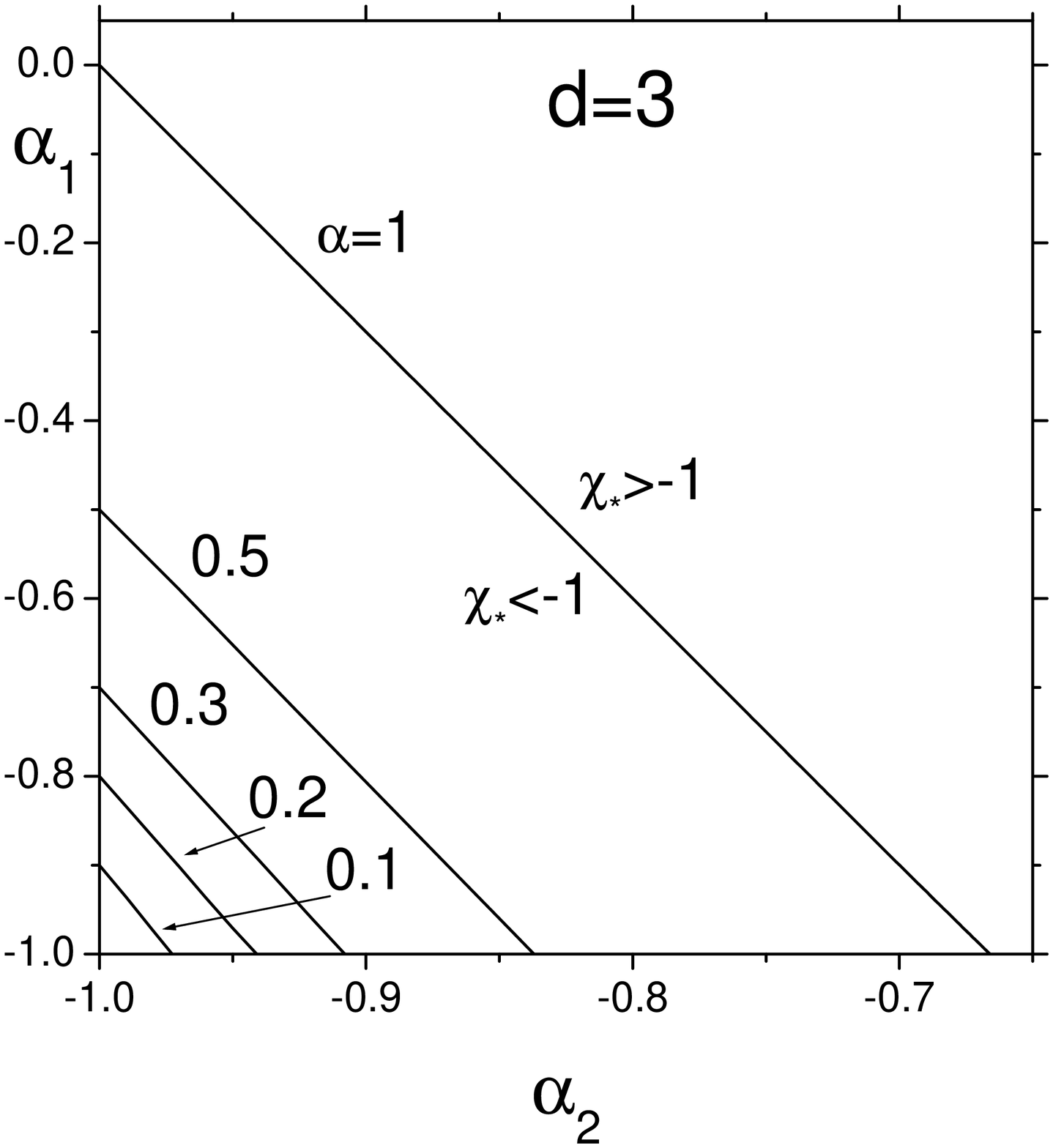}
   \end{flushright}
\vspace{-1.5cm} \caption{The restriction on the values of the
anisotropy parameters $\alpha_1$ and $\alpha_2$ for different values
of the compressibility parameter $\alpha$ for space dimensions $d=2$
and $d=3$ given by the physical condition $\chi_*>-1$. The figures
show that the allowed regions of the parameters $\alpha_1$ and
$\alpha_2$ decrease when the compressibility parameter $\alpha$
increases.  \label{fig1}}
\end{figure}

The IR scaling regimes of the model are given by the IR stable fixed
points of the corresponding RG equations
\cite{Vasiliev,AdAnVa96,AdAnVa99}. The fixed points of the RG
equations can be determined from the requirement  that all the
so-called beta functions of the model vanish and the IR stability of
the fixed point is given by the requirement that all the eigenvalues
of the matrix of the first derivatives $\Omega_{ij}=\partial
\beta_i/\partial C_k$, where $\beta_i$ denotes the full set of beta
functions and $C_k$ is the full set of charges of the model, must
have positive real parts. In our case the coordinates of the fixed
points are given by the following system of equations
\begin{equation}
\beta_g(g_*,\chi_{*},\alpha_j,\alpha,d,\varepsilon)=g_*
(-2\varepsilon+\gamma_{1}^*)=0,\quad
\beta_{\chi}(g_*,\chi_{*},\alpha_j,\alpha,d,\varepsilon)=\chi_*
(\gamma_1^*-\gamma_2^*)=0,\label{betaaaaa}
\end{equation}
for $j=1,2$ and variables with star denote the fixed point values of
the corresponding variables. The explicit form of the so-called
gamma functions $\gamma_i,i=1,2$ is as follows
\begin{equation}
\gamma_1=\frac{\bar{g}}{2} A_1\,, \quad \label{gammas}
\gamma_2=\frac{\bar{g}}{2\chi} A_2,
\end{equation}
where we denote $\bar{g}=g S_d/(2\pi)^d$,
$S_d=2\pi^{d/2}/\Gamma(d/2)$ is the $d-$dimensional sphere,  and
$A_{1,2}$ have the following explicit form
\begin{eqnarray}
A_1&=&\frac{(d+2)(d-1)+\alpha_1
(d+1)+\alpha_2+\alpha(d+2+\alpha_1-2\alpha_2)+\alpha^2\alpha_2}{d(d+2)},\label{AA1}
\\ A_2&=&
\frac{-2(\alpha_1+\alpha_2)+d^2\alpha_2+2\alpha(\alpha_1+d\alpha_2)+2\alpha^2\alpha_2}{d(d+2)}.\label{AA2}
\end{eqnarray}

\input epsf
   \begin{figure}[t]
     \vspace{-1cm}
       \begin{flushleft}
       \leavevmode
       \epsfxsize=7.5cm
       \epsffile{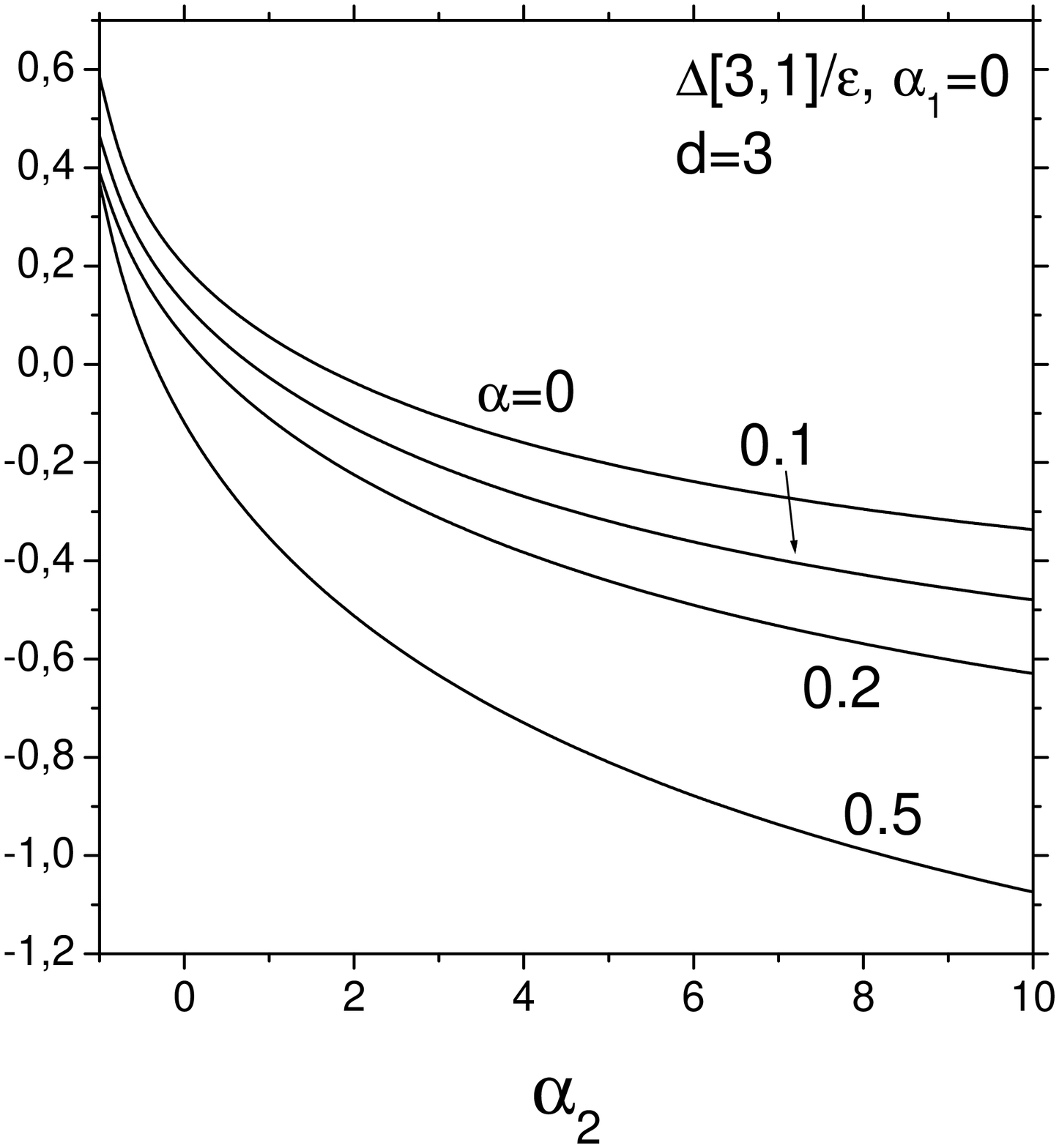}
   \end{flushleft}
     \vspace{-11.7cm}
   \begin{flushright}
       \leavevmode
       \epsfxsize=7.5cm
       \epsffile{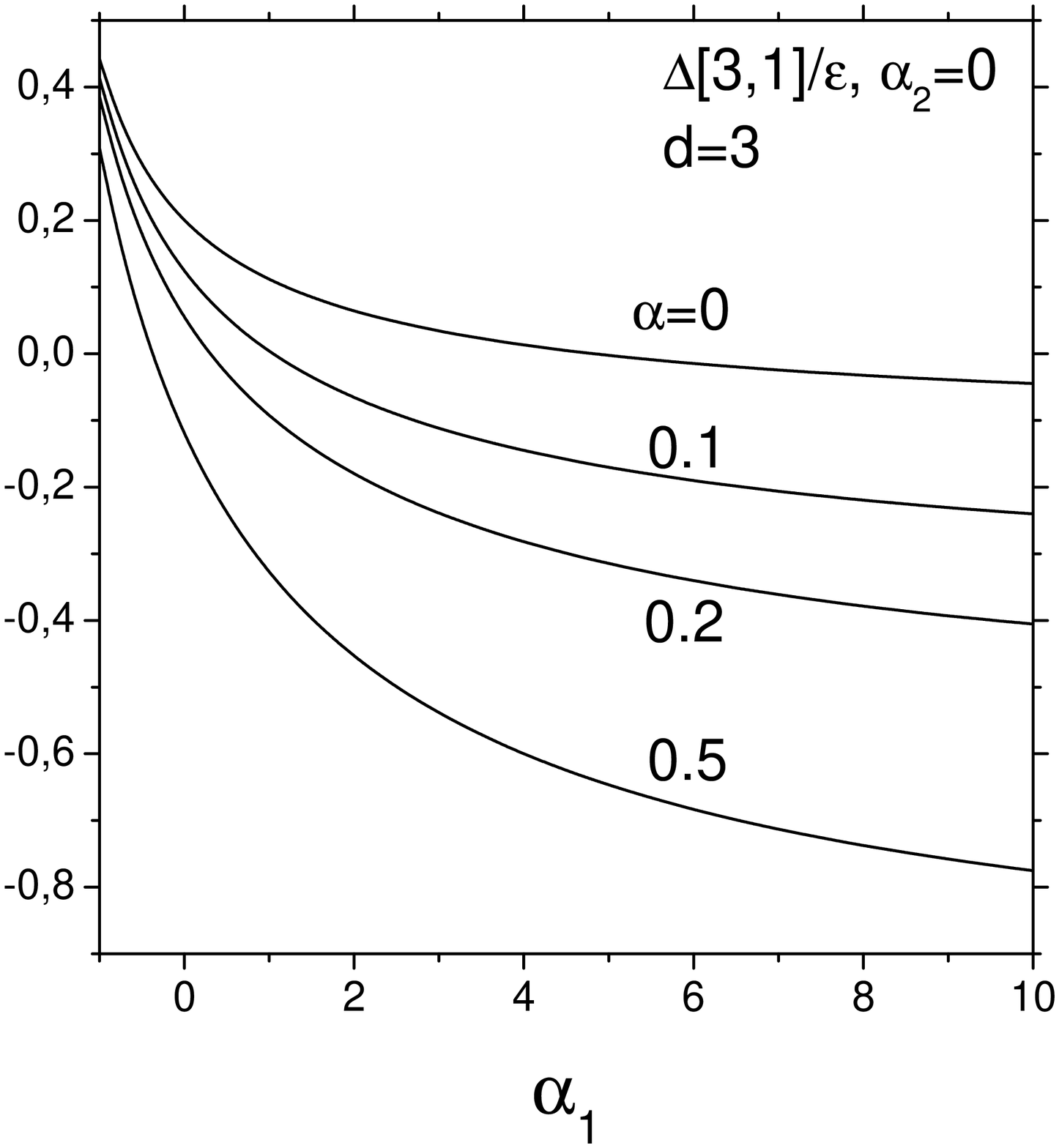}
   \end{flushright}
\vspace{-1.5cm} \caption{Dependence of the critical dimension
$\Delta[3,1]/\varepsilon$ on the anisotropy parameters $\alpha_1$
and $\alpha_2$ for different values of the compressibility parameter
$\alpha$ and for $d=3$.\label{fig2}}
\end{figure}

Thus, possible fixed points are found as solutions of the system of
algebraic equations (\ref{betaaaaa}) and their IR stability is
determined by the positive real parts of the eigenvalues of the
matrix $\Omega=\{\Omega_{ik}\}$. Using the explicit expressions for
gamma functions (\ref{gammas}) we have found the coordinates of the
nontrivial fixed point (trivial fixed point with $g_*=0$ is not
interesting for us)
\begin{eqnarray}
g_*&=&\frac{4 d (d+2)\varepsilon}{(d+2)(d-1)+\alpha_1
(d+1)+\alpha_2+\alpha(d+2+\alpha_1-2\alpha_2)+\alpha^2\alpha_2},\label{gAA1}
\\ \chi_*&=&
\frac{-2(\alpha_1+\alpha_2)+d^2\alpha_2+2\alpha(\alpha_1+d\alpha_2)+2\alpha^2\alpha_2}{(d+2)(d-1)+\alpha_1
(d+1)+\alpha_2+\alpha(d+2+\alpha_1-2\alpha_2)+\alpha^2\alpha_2}.\label{hAA2}
\end{eqnarray}
At fixed point the matrix of the first derivatives has the
eigenvalues $\lambda_{1,2}=2\varepsilon$. It means that the point is
IR stable if $\varepsilon>0$. It is also the only condition to have
$g_*>0$ (of course, together with the physical assumptions:
$\alpha_{1,2}>-1$ a $\alpha>0$). Besides, the fixed point value of
$\chi$ must satisfy physical condition $\chi_*>-1$. This fact leads
to restrictions on parametrical space of the model, namely, on
parameters of anisotropy $\alpha_{1,2}$ and compressibility
parameter $\alpha$. For space dimensions $d=2$ and $d=3$ these
restrictions are shown in Fig.\,\ref{fig1}. The existence of these
restrictions is not present in the model with simpler definition of
the anisotropy and compressibility which was studied in
\cite{JuJuReSc06a}.

Further we shall use the above obtained results to determine the
anomalous scaling of the single-time structure functions of the
scalar field inside the so-called inertial interval which is defined
by inequalities $1/\Lambda \ll r \ll L$
\cite{Frisch95,Vasiliev,AdAnHnNo00,JuJuReSc06a}. They are defined as
follows
\begin{equation}
S_{N}(r)\equiv\langle[\theta(t,{\bf x})-\theta(t,{\bf
x'})]^{N}\rangle, \label{struc}
\end{equation}
where $N$ denotes the order of the structure function and $r=|{\bf
r}|=|{\bf x}-{\bf x^{\prime}}|$. We shall not discuss details of the
corresponding analysis. It will be given elsewhere but we only
stress that the anomalous scaling is given by the existence in the
model of the so-called dangerous composite operator with negative
critical dimensions within the OPE
\cite{Vasiliev,AdAnVa96,AdAnVa99}. In our case the main contribution
to the anomalous scaling is given by the operators
$F[N,p]=\partial_{i_{1}}\theta\cdots\partial_{i_{p}}\theta\,(\partial_{i}\theta\partial_{i}\theta)^{n}$
with $N=2n+p$ (see, e.g., \cite{AdAnHnNo00,JuJuReSc06a}).

After rather long additional renormalization procedure
\cite{AdAnHnNo00,JuJuReSc06a} one comes to the following final
expression for the inertial range behavior of the single-time
structure functions
\begin{equation}
S_N({\bf r})=D_0^{-N/2} r^{N(1-\varepsilon)} \sum_{N^{\prime}\leq N}
\sum_p \{C_{N^{\prime,p}}\,(r/L)^{\Delta[N^{\prime},p]}+\dots\}\,,
\end{equation}
where $\Delta[N,p]$ denote the critical dimensions of the operators
$F[N,p]$ (more precisely, they are eigenvalues of the corresponding
matrix of the critical dimensions, see, e.g., \cite{JuJuReSc06a}),
$p$ obtains all possible values for given $N^{\prime}$,
$C_{N^{\prime,p}}$  are numerical coefficients which are functions
of the parameters of the model, and dots means contributions by the
operators others than $F[N,p]$ (see, e.g.,
\cite{Vasiliev,AdAnHnNo00} for details).

Our aim is to analyze the influence of compressibility on critical
dimensions of anisotropic operators and to find the answer on the
question whether the compressibility makes the anomalous behavior
more pronounced or not, i.e., whether the presence of
compressibility decreases the critical dimensions of corresponding
operators or not.

In Fig.\,\ref{fig2}, the dependence of the critical dimension of the
principal eigenvalue (related to the isotropic case, see
\cite{AdAnHnNo00}) of the corresponding matrix of critical
dimensions of the composite operators $F[3,p]$ is shown which play
the central role in the asymptotic behavior of the structure
function $S_3$ (it is well known that the structure function $S_2$
has not anomalous behavior \cite{AdAnHnNo00,JuJuReSc06a}, therefore
we start the analysis from $S_3$). It can be seen that in the case
of the structure function $S_3$ the anomalous scaling is present
(the existence of negative values for $\Delta[3,1]$) and the effect
is more pronounced for more compressible system. Detail analysis of
the model for the structure functions of higher order will be given
elsewhere.

\section*{Conclusions}

In present paper we have studied the influence of uniaxial
small-scale anisotropy  and compressibility on the anomalous scaling
of the structure functions of a scalar field in the framework of the
Kraichnan model by using the field theoretic RG and OPE. Using the
RG technique we have shown the existence of the scaling regime
within the inertial range which is defined by the corresponding IR
stable fixed point of the RG equations. We have found restrictions
on the anisotropy parameters which are related to the
compressibility of the system. The restrictions become larger when
compressibility parameter increases. Besides, using the OPE we have
found the asymptotic form of the structure functions and, using the
concrete example of the third order structure function, we have
shown that the compressibility of the system makes the effects of
anomalous scaling more pronounced, i.e., the critical dimensions of
the corresponding composite operators are smaller in compressible
case than in the incompressible case. Thus, we can conclude that the
compressible environment is more suitable for experimental study of
the anomalous scaling of the structure functions of a passive scalar
than incompressible systems.

\bigskip

\noindent  ACKNOWLEDGEMENTS --- It is a pleasure to thank the
Organizing Committee of the STM-2006 for kind hospitality. The work
was supported in part by VEGA grant 6193 of Slovak Academy of
Sciences, and by Science and Technology Assistance Agency under
contract No. APVT-51-027904.

\end{document}